\documentclass[a4paper,fleqn,usenatbib]{mnras}
\usepackage[T1]{fontenc}
\usepackage{ae,aecompl}

\usepackage{graphicx}	
\usepackage{amssymb}	

\usepackage[usenames]{color}

\newcommand {\bc}{\begin{center}}
\newcommand {\ec}{\end{center}}
\newcommand {\be}{\begin{equation}}
\newcommand {\ee}{\end{equation}}
\newcommand {\beq}{\begin{eqnarray}}
\newcommand {\eeq}{\end{eqnarray}}

\newcommand {\ergs}{{\rm erg\ \rm s^{-1}}}

\renewcommand{\d}{{\rm d}}

\title[Correlation between the cyclotron energy and luminosity]
{Positive correlation between the cyclotron line energy and luminosity in sub-critical X-ray pulsars: 
Doppler effect in the accretion channel}
\author[A. A.~Mushtukov et al.] 
{Alexander~A.~Mushtukov,$^{1,2}$\thanks{E-mail: al.mushtukov@gmail.com (AAM)}  
Sergey~S.~Tsygankov,$^{1}$ 
Alexander~V.~Serber,$^{3}$ 
\newauthor Valery~F.~Suleimanov $^{4,5}$ and
Juri Poutanen$^{1,6}$ \\ 
$^1$Tuorla observatory, Department of Physics and Astronomy, University of Turku,
  V\"ais\"al\"antie 20, FI-21500 Piikki\"o, Finland \\
$^2$Pulkovo Observatory, Russian Academy of Sciences, Saint Petersburg 196140, Russia \\
$^3$Nizhny Novgorod Planetarium, Revolutsionnaya st., 20, Nizhny Novgorod 603002, Russia\\
$^4$Institut f\"ur Astronomie und Astrophysik, Universit\"at T\"ubingen, 
    Sand 1, D-72076 T\"ubingen, Germany \\
$^5$Kazan (Volga region) Federal University, Kremlevskaja str., 18, Kazan 420008, Russia\\
$^6$Nordita, KTH Royal Institute of Technology and Stockholm University, Roslagstullsbacken 23, SE-10691 Stockholm, Sweden} 
\date{Accepted 2015 September 18. Received 2015 September 17; in original form 2015 July 17}

\pubyear{2015}

\begin{document}
\label{firstpage}
\pagerange{\pageref{firstpage}--\pageref{lastpage}}
\maketitle

\begin{abstract}
Cyclotron resonance scattering features observed in the spectra of some X-ray pulsars show significant changes of the line centroid energy with the pulsar luminosity. 
Whereas for bright sources above the so called critical luminosity these variations are established to be connected with the appearance of the high accretion column above the neutron star surface, at low, sub-critical luminosities the nature of the variations (but with the opposite sign) has not been  discussed widely. 
We argue here that the cyclotron line is formed when the radiation from a hotspot propagates through the plasma falling with a mildly relativistic velocity onto the neutron star surface. 
The position of the cyclotron resonance is determined by the Doppler effect. 
The change of the cyclotron line position in the spectrum with luminosity is caused by variations of the velocity profile in the line-forming region affected by the radiation pressure force. 
The presented model has several characteristic features: (i) the line centroid energy is  positively correlated with the luminosity; (ii) the line width is positively correlated with the luminosity as well; (iii) the position and the width of the cyclotron absorption line are variable over the pulse phase; (iv) the line has a more complicated shape than widely used Lorentzian or Gaussian profiles; (v) the phase-resolved cyclotron line centroid energy and the width are negatively and positively correlated with the pulse intensity, respectively. 
The predictions of the proposed theory are compared with the variations of the cyclotron line parameters in the X-ray pulsar GX 304--1 over a wide range of sub-critical luminosities as seen by the \textit{INTEGRAL}  observatory.
\end{abstract}

\begin{keywords}
pulsars: general -- scattering -- stars: neutron -- X-rays: binaries
\end{keywords}

\section{Introduction}
\label{intro}

X-ray pulsars (XRPs) are neutron stars (NSs) in binary systems accreting matter usually from a massive optical companion. 
Because of the strong magnetic field $B\gtrsim 10^{12}\,{\rm G}$) the accreting plasma is channelled toward the NS magnetic poles, where the accretion flow heats the surface and most of the gravitational energy is released via hard X-ray emission. 
The magnetic field modifies the observed X-ray spectrum often manifesting itself as the line-like absorption features, the so-called cyclotron lines. 
Such cyclotron resonance scattering features, sometimes also with harmonics, are observed in the spectra of about two dozens of XRPs \citep{2002ApJ...580..394C,2005AstL...31..729F,2015A&ARv..23....2W}.

In some cases, the luminosity-related changes of the line energy are observed, suggesting that the configuration of the line-forming region depends on the accretion rate. 
The line energy has been reported to be positively (for relatively low-luminous sources, $L\lesssim 10^{37}\,\ergs$; see \citealt{2007A&A...465L..25S,2012A&A...542L..28K}) and negatively correlated with luminosity (for relatively high-luminous sources, $L\gtrsim 10^{37}\,\ergs$; see \citealt{1998AdSpR..22..987M,2006MNRAS.371...19T,2007AstL...33..368T,2010MNRAS.401.1628T}), as well as uncorrelated with it \citep{2013ApJ...764L..23C}.

\begin{figure*}
\centering 
\includegraphics[width=15.5cm]{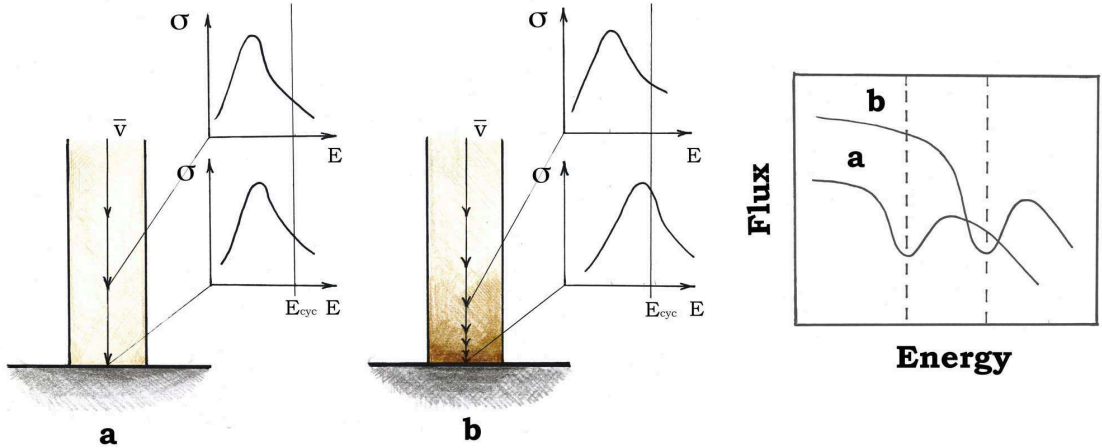}
\caption{The schematic presentation of the dependence of the cyclotron line energy on the velocity profile in the line-forming region.  
The radiation pressure affects the velocity profile when the luminosity is approaching its critical value: 
the higher the luminosity, the lower the velocity in the vicinity of NS surface. 
The lower the electron velocity, the lower the redshift and the higher the resonant energy at a given height. 
As a result, a positive correlation between the line centroid energy and the luminosity is expected. 
Panel (a) corresponds to the low-luminosity case when the gas is in a free-fall. 
The observed cyclotron line is redshifted  relative to the rest-frame value. 
Panel (b) is for the higher  luminosity when the gas is decelerated close to the surface and the line shifts closer to the rest-frame position. 
}
\label{pic:scheme}
\end{figure*}

The difference between positively and negatively correlated cases is explained by the absence or presence of the accretion column above the stellar surface, which begins to grow as soon as the accretion luminosity reaches its critical value  $L^* \sim 10^{37}\,\ergs$ \citep{1976MNRAS.175..395B,2015MNRAS.447.1847M,2015arXiv150603600M}. 
The negative correlation in a high-luminosity case was discussed in a number of works: \citet{2013ApJ...777..115P} proposed that the line is produced in the spectrum reflected from the NS surface and the anti-correlation with luminosity is explained by variations of the NS area illuminated by a growing accretion column. 
\citet{2014ApJ...781...30N} suggested that the growth of the column is associated with the decreasing magnetic field that leads to a shift in the cyclotron line position. 
This requires, however, that the column is only a few hundred meters high, contradicting the theoretical estimates giving a much taller column. 
The positive correlation in a low-luminous case has been explained by variations of the atmosphere height above the NS surface because of the changing ram pressure of the infalling material \citep{2007A&A...465L..25S}. 
This model is based, however, on an ad hoc assumption that the stopping depth of accreting protons (identified there with the atmosphere scale-height) is $\sim$100\,m instead of a more realistic value of 1\,m.
Alternatively, \citet{2014ApJ...781...30N} proposed that changes in the emission pattern towards more fan-like diagram at higher luminosity affect the position of the cyclotron line via the Doppler effect. 
There it was assumed that the velocity profile is not affected. 
This is, however, a gross simplification, because the accretion velocity in the vicinity of the NS surface is expected to change from the free-fall value of $\sim c/2$ to 0 with the grow of the luminosity  to the critical  value  $L^*$. 
Additionally, this model seems to contradict the positive correlation of the line width with luminosity \citep{2012A&A...542L..28K}, as 
the thermal width of the line is expected to be much smaller for directions across the magnetic field because of a weak transverse Doppler effect \citep[][]{Mushtukov2015PR}.

In this work we will focus only on the low-luminosity case (sub-critical XRPs). 
In this case, bulk of the radiation comes from the hotspot heated by the infalling plasma \citep{1969SvA....13..175Z,1993ApJ...418..874N}. 
This radiation, however, has to pass through the  material falling inside the accretion channel. 
The optical thickness of the channel at the resonant energies is much higher than unity in the considered luminosity range, $L\sim 10^{35} \div 10^{37}\,\ergs$. 
The photons are resonantly scattered there, leading to the formation of a cyclotron absorption-like feature in the spectrum. 
Plasma in the accretion flow moves with a significant velocity towards the NS surface, and the scattering feature must be redshifted due to the Doppler effect.
An observer thus sees the line at a lower energy than the cyclotron one associated with the magnetic field at the pole. 
The amplitude of the shift depends on the electron velocity and the angle between the photon momentum and the accretion flow velocity. 
As a result, the observed position and the shape of the cyclotron scattering feature depend on the velocity profile in the accretion channel (see Fig.~\ref{pic:scheme}). 
Particularly, it is expected that the lower the electron velocity in the region, the lower the redshift and the higher the line centroid energy. 
In this case, it is also closer to the actual cyclotron energy $E_{\rm cyc}\simeq 11.6 B_{12}\,{\rm keV}$, which is defined by the surface magnetic field strength $B_{12}\equiv B/10^{12}\,{\rm G}$. 
The radiation pressure affects the velocity profile when the luminosity is approaching its critical value \citep{1976MNRAS.175..395B,2015MNRAS.447.1847M}: the higher the luminosity, the lower the velocity in the vicinity of NS surface. 
Therefore, we expect a positive correlation between the line centroid energy and the XRP luminosity.

\begin{figure}
\centering 
\includegraphics[width=8.5cm]{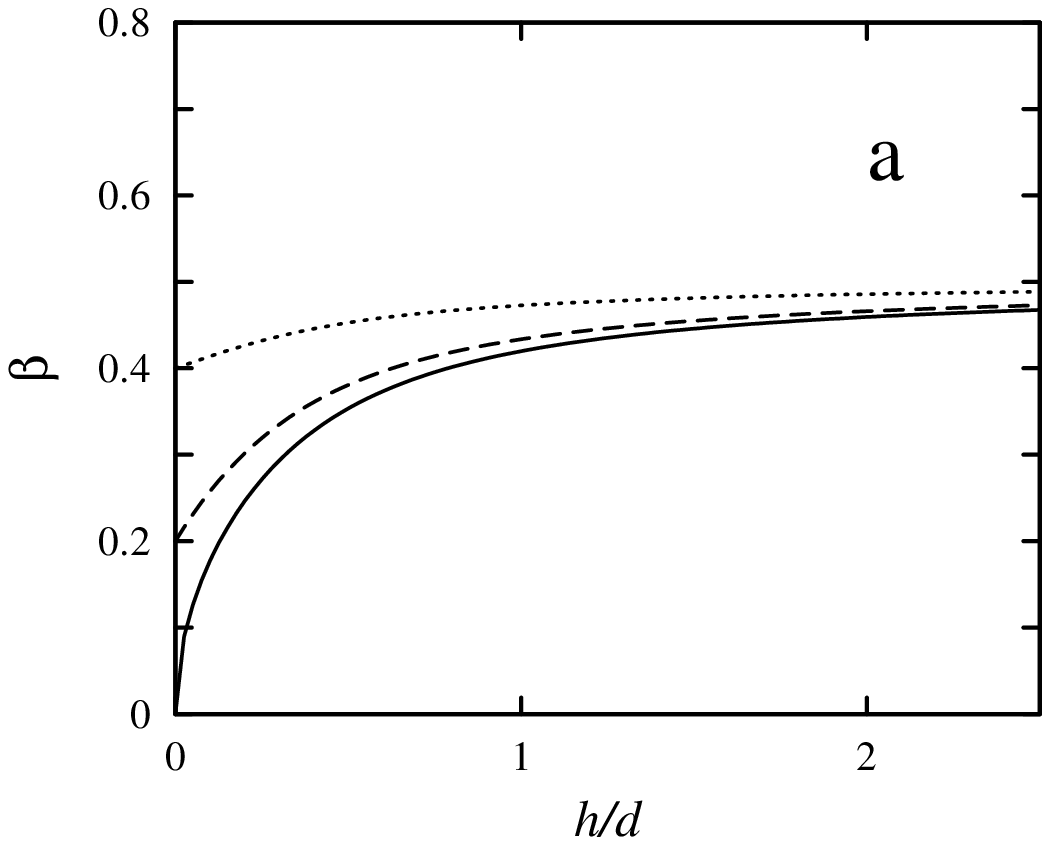} 
\includegraphics[width=8.5cm]{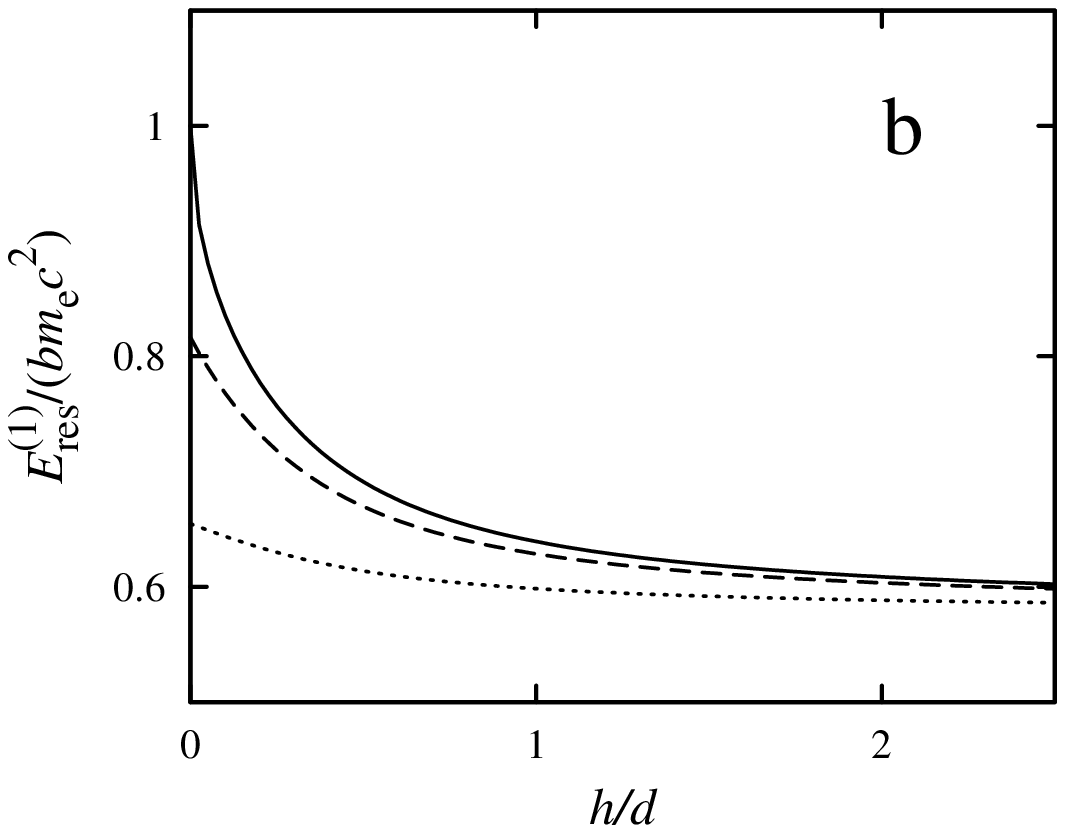} 
\caption{The dependence of (a) the dimensionless velocity  and  (b) the corresponding resonant scattering energy in the laboratory reference frame on the distance from the NS surface measured in units of the thickness of the accretion channel. 
The velocity profile is given by equations~(\ref{eq:vel_prof}) and (\ref{eq:vel_prof_norm}) and  the cyclotron energy is computed from equation~(\ref{eq:res_energy}). 
Different curves correspond to various final electron velocities: $\beta_{0}=0$ (solid), $\beta_{0}=0.2$ (dashed), $\beta_{0}=0.4$ (dotted). 
Here, the free-fall velocity is $\beta_{\rm ff}=0.5$ and the observing angle is $\theta=0$.
}
\label{pic:vel_ene}
\end{figure}

In this paper, we construct a simple model, which describes variations of the opacity near the cyclotron energy, and compare the predictions of our model with the observations obtained by the \textit{INTEGRAL} space observatory of the XRP GX 304--1 during its Type~II outburst in the beginning of 2011.

\section{The model}
\label{BasicIdeas}

The basic process which determines the interaction between the radiation and matter in the vicinity of a NS surface is Compton scattering. 
In case of strong magnetic field the scattering has a number of special features. Its cross section strongly depends on the $B$-field strength, photon energy, polarization and momentum. 
Electron transitions between Landau levels cause the resonance scattering at some photon energies, where the cross section can exceed the Thomson scattering cross section by several orders of magnitude. 
The resonant scattering leads to appearance of cyclotron absorption features in the spectra of XRPs. 
The exact shape and position of the features in the spectra are determined by the structure of the line-forming region.

The exact resonant energy depends on the velocity of the electrons because of  the effects of special relativity. 
In a particular case of cold electrons in bulk motion with the velocity $\beta\equiv v/c$ towards the NS surface, the position of the fundamental is \citep{1986ApJ...309..362D,M1992,2006RPPh...69.2631H}
\be \label{eq:res_energy}
\frac{E^{(1)}_{\rm res}}{m_{\rm e}c^2}=\frac{\sqrt{[\gamma(1 -\beta \cos\theta')]^2+2b\sin^2\theta'}-\gamma(1 -\beta \cos\theta')}
{\sin^2\theta'},
\ee
where  $\gamma\equiv(1-\beta^2)^{-1/2}$ is the Lorentz factor, $\theta'$ is the angle between the electron and photon momentum, $b=B/B_{\rm cr}$ is the local magnetic field strength in units of the critical magnetic field $B_{\rm cr}=m^2_{\rm e}c^3/(e\hbar)=4.413\times 10^{13}\,{\rm G}$. 
For $b\ll 1$ or $\theta'\ll 1$, the expression for the fundamental in the observer (laboratory) reference frames is reduced to 
\be\label{eq:DoppShift}
\frac{E^{(1)}_{\rm res}}{m_{\rm e}c^2}\simeq \frac{b}{ \gamma (1+\beta\cos\theta)}, 
\ee
where $\theta=\pi-\theta'$ is the angle between the line of sight and the magnetic axis (normal to the NS surface).
We see that the resonant energy in the laboratory reference frame is always redshifted and the redshift depends on the plasma velocity and the observing angle.

The velocity profile of the infalling material in the vicinity of the NS surface is determined by the radiation flux $F$ and the effective cross-section $\sigma$, because the accreting matter is interacting with radiation and is decelerating to some final velocity $\beta_0=v_0/c$ at the NS surface. 
Then it looses its momentum entirely due to Coulomb collisions \citep{1969SvA....13..175Z}. 
The velocity just above the NS surface $\beta_0$ and the velocity profile in the accretion channel depend on the radiation pressure force $F\sigma/c$ and therefore on the luminosity of an XRP. 
Obviously, the velocity $\beta_0$ tends to be smaller for a higher luminosity and it turns to zero when the XRP luminosity reaches its critical value 
\be 
L^*=4\times 10^{36}\,\frac{m}{R_6}\left(\frac{l}{2\times 10^5 {\rm cm}}\right)\frac{\sigma_{\rm T}}{\sigma}\,\,\ergs,
\ee 
where $m=M/{\rm M}_\odot$ is the NS mass in units of the solar mass, $R=10^6 R_6$~cm is the NS radius, $l$ is the length of the accretion channel footprint, which has a shape of thin annual arc, and $\sigma_{\rm T}$ is the Thomson  cross-section. The critical luminosity $L^*$ strongly depends on the $B$-field strength \citep{1976MNRAS.175..395B,2015MNRAS.447.1847M}.  At higher luminosity  the accretion column begins to grow above the surface.

If the luminosity $L\sim L^*$, the gravitational force is much lower than the radiation pressure force, which defines the deceleration of matter. 
In order to obtain the velocity profile, we need to account for the dependence of the radiation flux on the height $h$ above the NS surface. 
In case of a filled accretion funnel, the dependence is
\be
F(h)\simeq F_0 \frac{1}{1+(h/d)^2},
\ee
where $F_0$ is the flux at the surface and $d$ here is the radius of the hot spot. 
Then the velocity changes with height according to the equation
\be
\frac{\d \beta^2}{\d h}=\frac{2\sigma F(h)}{m_{\rm p}c^3},
\ee
where $m_{\rm p}$ is the proton mass. 
Thus we get (see Fig.\,\ref{pic:vel_ene}) 
\beq\label{eq:vel_prof}
\beta(h)=\left[\beta^2_{\rm ff}-\frac{\pi \sigma F_{0}d}{m_{\rm p}c^3}\left(1-\frac{2}{\pi}\tan^{-1}(h/d)\right)\right]^{1/2}, 
\eeq
where 
\be\label{eq:vel_prof_norm}
\frac{\pi \sigma F_{0} d}{m_{\rm p}c^3} =  \beta^2_{\rm ff}-\beta^2_0.
\ee 
If $\beta_0=0$, then $F_0$ becomes the critical flux $F^*$ corresponding to the critical luminosity $L^*=2\pi d^2F^*$. 
It is also obvious that the luminosity corresponding to a given velocity $\beta_0$  at the bottom of the accretion channel is
\be \label{eq:L_general}
L(\beta_0) \simeq L^* \left( 1- \frac{\beta_0^2}{\beta_{\rm ff}^2} \right) . 
\ee

If the accreting matter is confined to a narrow wall of the magnetic funnel, the flux dependence on the height is more complicated. The flux drops as  $(1+h^2/d^2)^{-1/2}$ when $h\ll l/(2\pi)$ and as $(1+4\pi^2 h^2/l^2)^{-1}$ when $h\gtrsim l/(2\pi)$, resulting in a slightly different velocity profile. 
However, the qualitative behaviour is the same (see also \citealt{1982SvAL....8..330L,1998ApJ...498..790B}) and further in this paper we use equation (\ref{eq:vel_prof}) for calculation of the velocity profile.

Radiation from the hot NS surface has to pass through the layer of moving plasma and is resonantly scattered there by electrons. 
The resonant energy depends on the local electron velocity (see Fig.\,\ref{pic:vel_ene}). 
As a result, it is expected that the position of the absorption line feature in the spectrum and its shape are determined by the plasma velocity profile near the NS surface. 
Higher luminosity would lead to lower velocity shifting the resonance to higher energy in the observer reference frame.

\subsection{The opacity of moving plasma near the resonance energy}
\label{sec:TheOpacityCalc}

In order to define main characteristics of the cyclotron line and its variability with changes of the accretion luminosity, we have to calculate the opacity of the line-forming region as a function of photon energy $E$.
The opacity in a given direction (defined by the zenith angle $\theta$ and the azimuthal angle $\phi$) for photons of energy $E$ emitted at the distance $x$ from the accretion channel edge is  
\be
\tau(E,\theta,\phi,x)\simeq\int\limits_{0}^{y^*(\theta,\phi)}\frac{\d y}{\cos\theta}\,\sigma\left(E,\theta,y\right)n_{\rm e}(y),
\ee
where $y^{*}(\theta,\phi)\simeq x/(\tan\theta\cos\phi)$ for the case of $l\gg d$ (in our case $l/d\simeq 160\,L^{-4/35}_{37}B^{-1/14}_{12}m^{71/140}$, $L_{37}\equiv L/10^{37}\ergs$, see \citealt{2015arXiv150603600M}), $\sigma(E,\theta,y)$ is the scattering cross section at height $y$ above the stellar surface, $n_{\rm e}(y)$ is the local electron number density. 
The scattering cross section $\sigma\left(E,\theta,y\right)$ depends on the local temperature $T(y)$ as well. 
We use the scattering cross-section computed using second order perturbation theory of quantum-electrodynamics \citep{1986ApJ...309..362D,1991ApJ...374..687H,2012PhRvD..85j3002M,Mushtukov2015PR}. 

The resonant scattering cross-section of the X-mode photons is higher than the resonant scattering cross-section of the O-mode photons \citep{1971PhRvD...3.2303C,1981ApJ...251..288N,Mushtukov2015PR}. 
As a result, the emission around the cyclotron line should be dominated by the O-mode photons. 
Therefore,  we use the O-mode cross-sections in the following calculations. 
We also assume that the electron temperature of the accreting gas is constant and we fixed it at $3\,{\rm keV}$, a value close to the Compton temperature of radiation. The results are not affected by the specific choice of that temperature.
Then the average value of the opacity over the coordinate $x$ inside the accretion channel and the azimuthal angle  is
\be\label{eq:tau2}
\overline{\tau}(E,\theta)\simeq\frac{2}{d\pi} \int\limits_{0}^{d}\d x\int\limits_{0}^{\pi/2} \d\phi\,\tau(E,\theta,\phi,x) \exp(-\tau_{\rm T}(\theta,\phi,x)),
\ee
where  $\tau_{\rm T}(\theta,\phi,x)$ is the Thomson optical thickness at the direction given by angles $\theta$ and $\phi$, which affects the residual continuum intensity at a given direction from given point: $I(\theta,\phi)\approx I(\theta,\phi,x)\exp(-\tau_{\rm T}(\theta,\phi,x))$. 
If the accretion disc is interrupted in the C-zone \citep{SS1973,Sul2007}, where the gas pressure and the Kramer opacity dominate,  the accretion channel thickness can be estimated as follows:
\be
d\approx 6\times 10^3\, \Lambda^{-3/8}L^{9/35}_{37}B^{-3/14}_{12}m^{-81/140}R^{39/35}_{6}\, \mbox{cm}, 
\ee
with $\Lambda=0.5$ being a commonly used value \citep{GL1978}. 

In order to compare with observations, we then compute the  opacity averaged over the pulsar phase $\varphi$:
\be\label{eq:tau_a1a2}
\overline{\tau}(E)\simeq\frac{\int_{0}^{\pi}\overline{\tau}(E,\theta)|\cos\theta|\,\d\varphi}{\int_{0}^{\pi}|\cos\theta|\d\varphi},
\ee
where $\cos\theta=\cos i\cos\alpha+\sin i\sin\alpha\cos\varphi$, $i$ is the angle between the line of sight and the NS rotational axis, $\alpha$ is the magnetic inclination and $\varphi$ is the phase angle. 
Here we assumed a blackbody-like emission pattern with the emitted flux proportional to the cosine of the hotspot inclination to the line of sight $\cos\theta$.

Assuming dipole magnetic field structure, the field strength in the vicinity of stellar surface depends on height $h$ as follows: $B(h)=B_0 \left(\frac{R}{R+h}\right)^3$, where $B_0$ is the surface magnetic field strength. 
From the mass conservation law $\dot{M}/(2S_{\rm D})\simeq m_{\rm p}n_{\rm e}(h)v(h)$ we can get the local electron number density:
\be
n_{\rm e}(h)\approx 2\times 10^{30}\frac{\dot{M}_{17}}{2\,S_{\rm D}\beta(h)}\approx 2\times 10^{19}\,L^{3/5}_{37}B^{-1/2}_{12}\beta^{-1}(h) \, \mbox{cm}^{-3} , 
\ee
where $\beta(h)=v(h)/c$ is dimensionless velocity and $S_{\rm D}$ is the hotspot area:
\be
S_{\rm D}\approx 3 \times 10^9\, \Lambda^{-7/8}\, m^{-13/20}\, R_6^{19/10}\, B_{12}^{-1/2}\, L_{37}^{2/5}\quad\mbox{cm}^2. 
\ee
Thus for a given luminosity, we compute the velocity profile in the line-forming region,  which affects the opacity, and then for given parameters $i$ and $\alpha$ we can compute the average opacity using equation~(\ref{eq:tau_a1a2}).

\subsection{Opacity profiles}
\label{sec:TheMainFeatures}

\begin{figure}
\centering 
\includegraphics[width=8.5cm]{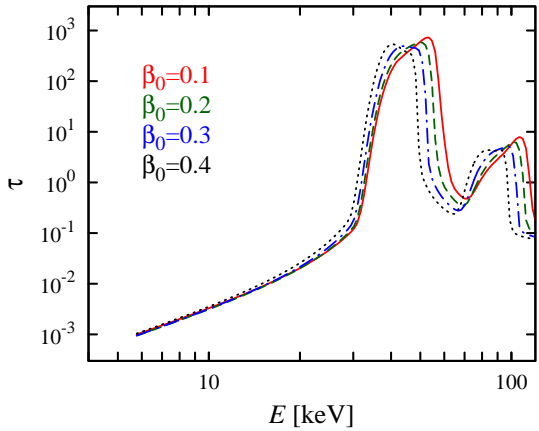} 
\caption{The averaged opacity profile for various final velocities $\beta_0$ at the NS surface. 
The profiles are computed for the fixed accretion luminosity $L=10^{37}\,\ergs$. 
We see that the lower the final velocity, the higher the centroid energy and the wider the feature. 
Here $B_0=5\times 10^{12}\,{\rm G}$, $i=60^{\circ}$, $\alpha=20^{\circ}$.}
\label{pic:opas_v}
\end{figure}

For our calculations we have taken a NS of mass $M=1.4\,{\rm M}_{\odot}$, radius $R=10\,{\rm km}$ and surface dipole magnetic field strength at the poles $B_0=5\times 10^{12}\,{\rm G}$. 
The opacity averaged over the orientation of line-forming region (see eq.\,(\ref{eq:tau_a1a2})) as a function of photon energy is presented in Fig.\,\ref{pic:opas_v}, where different lines correspond to different final velocity on the NS surface. 
The peak of the opacity, whose complicated shape is defined by the velocity distribution in the line-forming region, shifts to higher energies when the surface velocity of the accretion flow decreases. 
The lower the surface velocity, the wider the opacity peak produced by resonant scattering. 
The opacity profile determines the shape of the scattering feature in the XRP spectrum. 
Therefore, taking into account the connection between the accretion luminosity and the velocity $\beta_0$ we find that the cyclotron absorption feature centroid energy and its width should be positively correlated with the luminosity for sub-critical XRPs: the higher the luminosity, the higher the line centroid energy, the wider the cyclotron line in the spectrum. 
The increase of the centroid energy is caused by the appearance of a layer with a lower velocity, where the redshift of the cyclotron energy is lower. 
This fact explains also the increase of the line width: it grows on its blue side because of the layers of lower velocity. 
We also note that the shape of the line is asymmetric and can hardly be approximated by the widely used Lorentzian or Gaussian profiles. 

\begin{figure}
\centering 
\includegraphics[width=8.5cm]{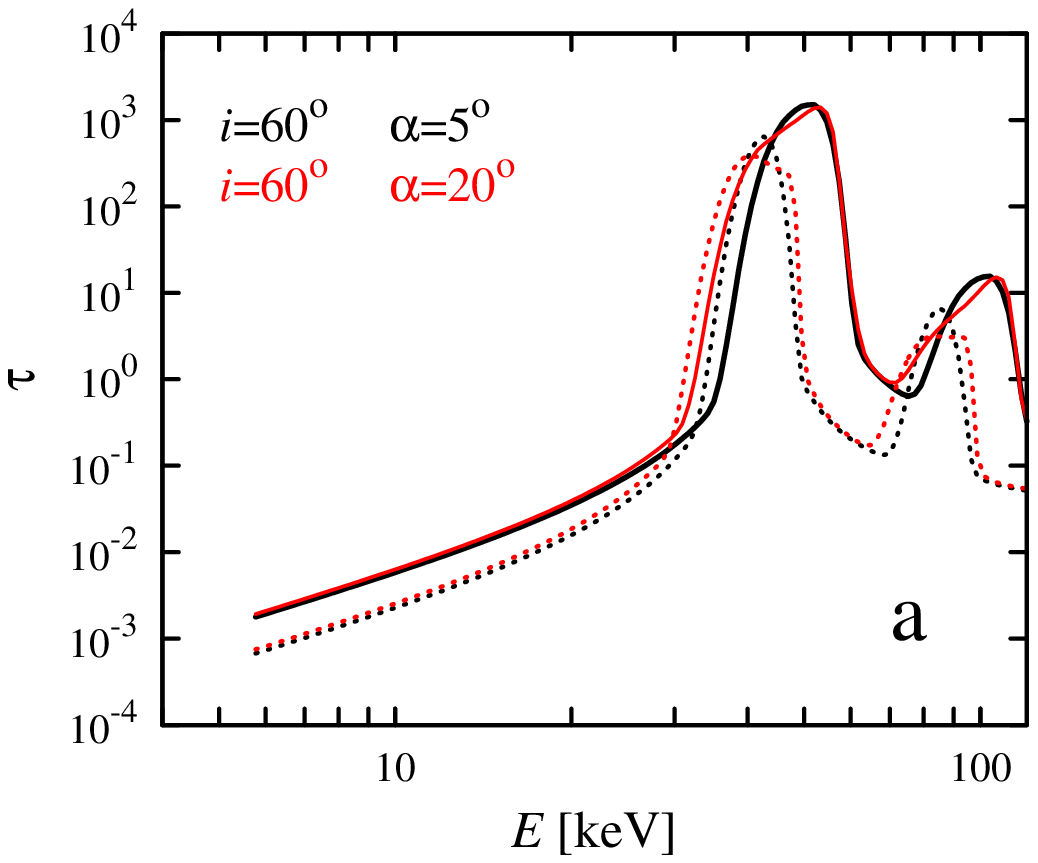} 
\includegraphics[width=8.5cm]{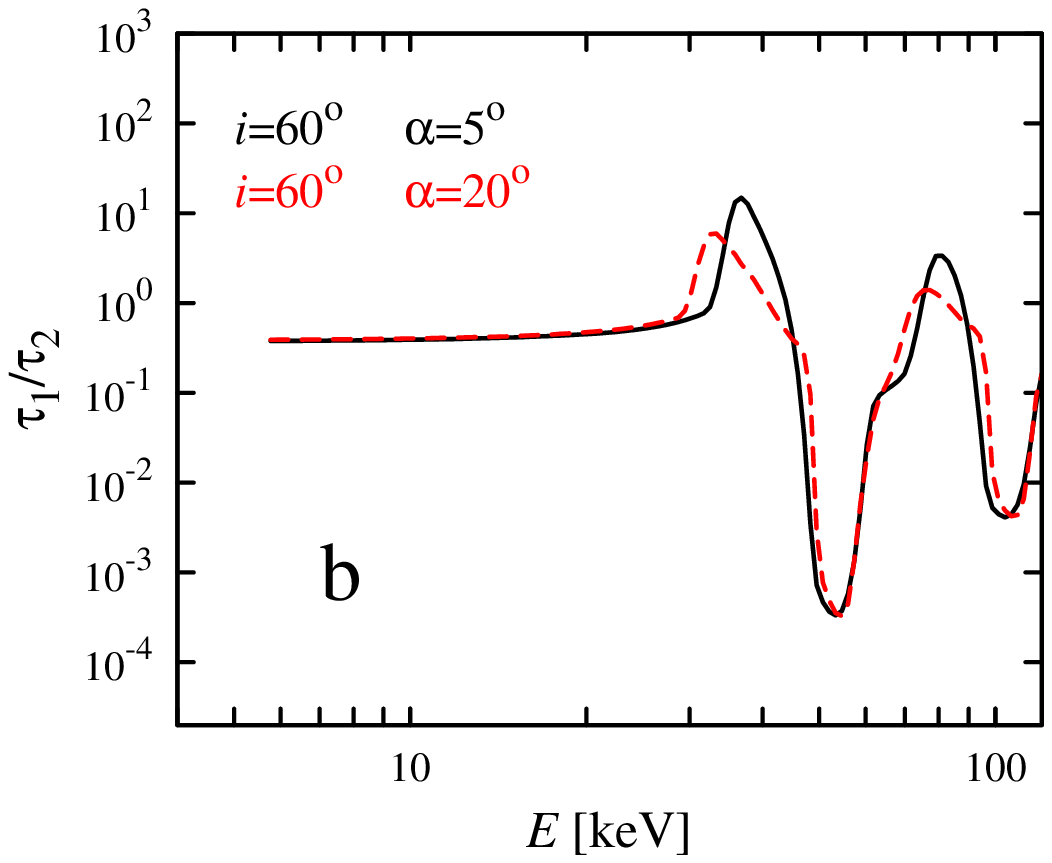} 
\caption{(a) The opacity profile averaged over the pulse period for various magnetic inclination $\alpha$ (see eq.~(\ref{eq:tau_a1a2})). Black lines are for $i=60^\circ$ and $\alpha=5^\circ$, while red lines are for $i=60^\circ$ and $\alpha=20^\circ$. Solid and dotted lines are given for $\beta_0=0.1$ and $\beta_0=0.4$, respectively. 
(b) The ratio of the opacity profiles $\tau_1=\tau(\beta_0=0.4)$, $\tau_2=\tau(\beta_0=0.1)$ corresponding to the different velocities at the NS surface. 
Here $B=5\times 10^{12}$~G.
}
\label{pic:opas_ration}
\end{figure}

The behaviour of the line can be well illustrated by the ratio of opacities calculated for different surface velocity values (see Fig.\,\ref{pic:opas_ration}). 
If one takes the ratio of the low-luminosity opacity (high $\beta_0$) to the high-luminosity opacity (low $\beta_0$), then the  shape of the ratio is quite conservative and does not depend on the angles, which the opacity is averaged over: the ratio reaches a local maximum and then has a deep local minimum (see Fig.\,\ref{pic:opas_ration}b). 
The same behaviour is expected for the ratio of higher luminosity XRP spectrum to the  lower luminosity spectrum.  
Because we do not have a model for the spectrum formation, the analysis of spectral ratios can be suggested as a good test for the model.

It is also expected that the cyclotron line absorption feature has to show variability during the pulse period. 
The Doppler shift of the resonant energy is lower for larger angles between the accretion flow velocity and the line of sight (see eq.~(\ref{eq:DoppShift})). 
Therefore, it is expected that the phase-resolved line centroid energy would be anti-correlated with the pulse intensity for sub-critical XRP with a pencil beam emission diagram \citep{1973A&A....25..233G}. 
Because the resonance peak in the scattering cross-section tends to be wider for the photons propagating along the $B$-field lines \citep{1991ApJ...374..687H,Mushtukov2015PR}, it is expected that the phase-resolved cyclotron line width would be positively correlated with the pulse intensity. 
The radiation beam pattern also affects  the phase-averaged spectrum and its changes with the luminosity can affect the cyclotron line energy \citep{2014ApJ...781...30N}.
In our calculation the radiation beam pattern is taken into account in equation\,(\ref{eq:tau2}), where the opacity is weighted with the continuum intensity escaping in different direction.


\section{Comparison with observations}
\label{sec:observ}

\subsection{Observational data}

The most reliable positive correlation between the cyclotron line energy and the luminosity has been observed in the spectra of XRP GX~304--1 \citep{2011PASJ...63S.751Y,2012A&A...542L..28K}. 
Therefore, we use this source as a case study to verify the predictions of our model.

GX~304--1 belonging to the subclass of XRPs with Be-type optical companions \citep{1978MNRAS.184P..45M} has a relatively long pulse period of 272~s \citep{1977ApJ...216L..15M}. 
It is located at a distance of about 2.4~kpc \citep{1980MNRAS.190..537P}. 
The cyclotron absorption line at 54~keV in the source energy spectrum was discovered only recently \citep{2011PASJ...63S.751Y} using \textit{Suzaku} and \textit{RXTE} data obtained during Type~I outbursts recurrently observed from this system every 132.5~d.

In our current work, we reanalyse the data obtained by the \textit{INTEGRAL} observatory \citep{2003A&A...411L...1W} during another Type~II outburst occurred in the beginning of 2011 \citep{2012ATel.3856....1Y}. 
The data sample is identical to the one utilized by \cite{2012A&A...542L..28K}. 
The total duration of the outburst was about one month. 
It was evenly covered by eight \textit{INTEGRAL} observations (during spacecraft revolutions 1131--1138). 
Both rising and declining parts as well as the maximum of the outburst were monitored.

\begin{table*}
\caption{Cyclotron line parameters for the \textit{INTEGRAL} observations analysed in this work.}\label{tab:observ}
  \label{multiprogram}
\centering
  \begin{tabular}{|c|c|c|c|c|c|c|c|}
    \hline
 Revolution & $L_x$ 	$^a$        & \multicolumn{3}{|c|}{CYCLABS} &  \multicolumn{3}{|c|}{GABS} \\
    \cline{3-8}
number      & ($10^{37}$ erg s$^{-1}$) & $E_{\rm cyc}$ & $\sigma_{\rm cyc}$ & $\tau_{\rm cyc}$ & $E_{\rm cyc}$ & $\sigma_{\rm cyc}$ & $\tau_{\rm cyc}$ \\ 
& & (keV) & (keV) & & (keV)  &(keV)  &  \\
    \hline
1131 & 0.50 & $54.1^{+0.8}_{-0.8}$ & $9.0^{+1.7}_{-1.5}$ & $0.65^{+0.07}_{-0.07}$ & $56.3^{+1.2}_{-1.1}$ & $7.5^{+0.9}_{-0.8}$ & $10.4^{+2.0}_{-1.8}$ \\
1132 & 1.12 & $55.9^{+0.5}_{-0.5}$ & $11.8^{+1.3}_{-1.2}$ & $0.74^{+0.05}_{-0.05}$ & $59.2^{+0.9}_{-0.8}$ & $9.1^{+0.7}_{-0.7}$ & $14.3^{+2.1}_{-1.9}$\\
1134 & 2.20 & $56.1^{+0.5}_{-0.5}$ & $14.8^{+1.8}_{-1.6}$ & $0.68^{+0.06}_{-0.05}$ & $61.5^{+0.8}_{-0.7}$ & $11.7^{+0.7}_{-1.1}$ & $18.8^{+1.5}_{-1.2}$\\
1136 & 0.96 & $52.6^{+0.6}_{-0.6}$ & $7.7^{+1.4}_{-1.3}$ & $0.53^{+0.05}_{-0.05}$ & $54.5^{+0.8}_{-0.7}$ & $6.8^{+0.7}_{-0.7}$ & $7.5^{+1.3}_{-1.1}$\\
1137 & 0.47 & $53.1^{+0.8}_{-0.7}$ & $5.1^{+1.5}_{-1.4}$ & $0.48^{+0.08}_{-0.07}$ & $54.3^{+0.9}_{-0.9}$ & $5.9^{+0.8}_{-0.8}$ & $5.5^{+1.2}_{-1.0}$\\
1138 & 0.22 & $49.6^{+1.6}_{-1.5}$ & $3.9^{+2.5}_{-2.7}$ & $0.39^{+0.56}_{-0.12}$ & $50.6^{+1.9}_{-1.6}$ & $4.8^{+1.5}_{-1.3}$ & $3.8^{+1.7}_{-1.3}$\\
    \hline
  \end{tabular}
  \begin{flushleft}{ 
$^{a}$ Luminosity in the 3--100 keV energy band, assuming source distance of $2.4$ kpc.   }\end{flushleft} 
\end{table*}

The data from the IBIS/ISGRI detector  \citep{2003A&A...411L.141L} were screened and reduced following the procedures described in \cite{2010A&A...519A.107K} and \cite{2014Natur.512..406C}. 
Energy calibration was provided by the Offline Scientific Analysis (OSA) version 10.1 \citep{2013arXiv1304.1349C}. 
Additional gain correction based on the position of the tungsten fluorescent line (at 58.2~keV) was applied, keeping the uncertainty of energy reconstruction to less than 0.1 keV. 
Data from the  JEM-X monitor \citep{2003A&A...411L.231L} were processed with the standard procedures from OSA 10.1. 

\begin{figure}
\includegraphics[width=\columnwidth,bb=55 150 550 710, clip]{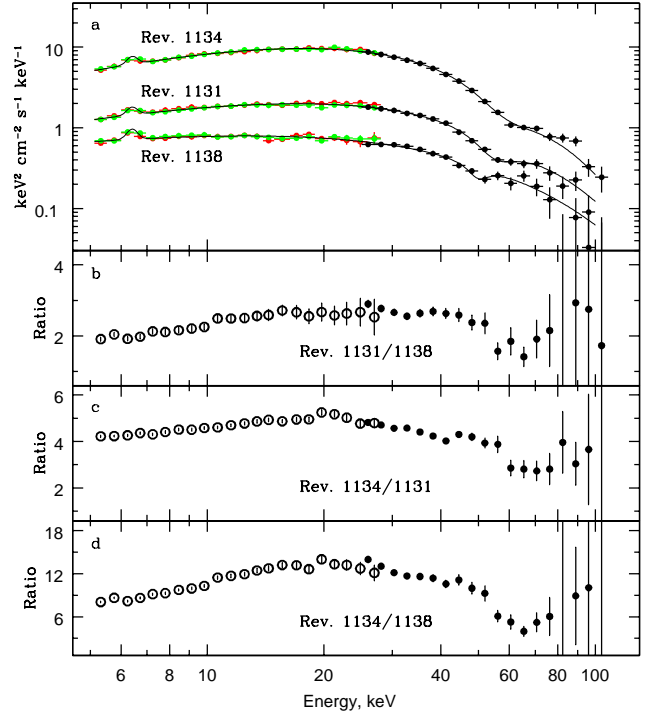}
\caption{(a) Energy spectra of GX 304--1 obtained by main instruments
  of the \textit{INTEGRAL} observatory during different spacecraft
  revolutions. Data in the energy range 5--30 keV are from
  {JEM-X} monitor (red and green points for {JEM-X1} and
  {JEM-X2} modules), between 25 and 100 keV -- from the
  {IBIS} telescope. 
  (b)--(d) Ratio of spectra obtained during different revolutions. 
  }\label{pic:specs}

\end{figure}

Observed fluxes from GX~304--1  range between about 0.1 and 1 Crab, permitting us to trace the evolution of the energy spectrum in a wide energy range (5--100 keV) over very different luminosities of the source. 
Three spectra obtained at different states are shown in Fig.~\ref{pic:specs}(a). 
Particularly, spectra obtained during the lowest-luminosity state at the rising phase, peak phase, and lowest-luminosity state at the declining phase of the outburst are denoted by Rev. 1131, Rev. 1134, and Rev. 1138, respectively. 
JEM-X data are used in the range 5--30 keV and shown by  green and red circles (for {JEM-X1} and JEM-X2 modules, correspondingly),  {IBIS/ISGRI} data cover the energy range from 25 to 100 keV (shown by filled black circles). 
The solid curves represent the best-fitting models that consist of a power law with an exponential cutoff modified by the Gaussian line at 6.4 keV and the cyclotron absorption feature in the form of Lorentzian optical depth profile ({\sc cyclabs} model in {\sc XSPEC}; \citealt{1990Natur.346..250M}).
The final form of the fitting model in {\sc XSPEC} package can be expressed as {\sc (gau+powerlaw*highecut)*cyclabs}.

\begin{figure}
\includegraphics[width=8.5cm]{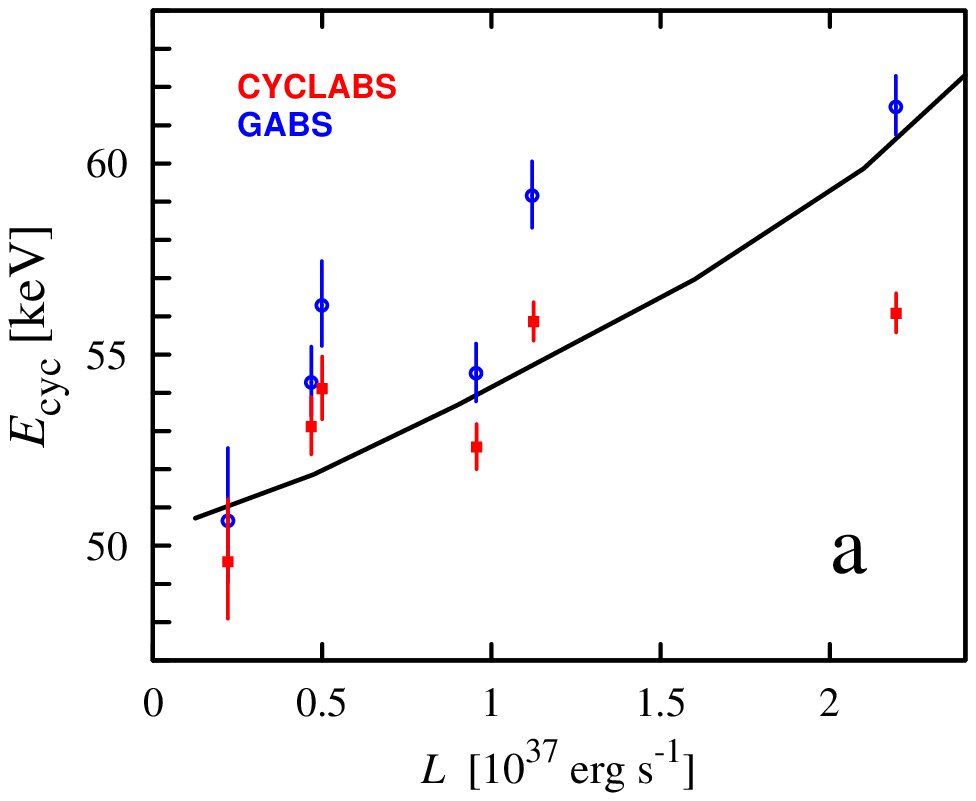} 
\includegraphics[width=8.5cm]{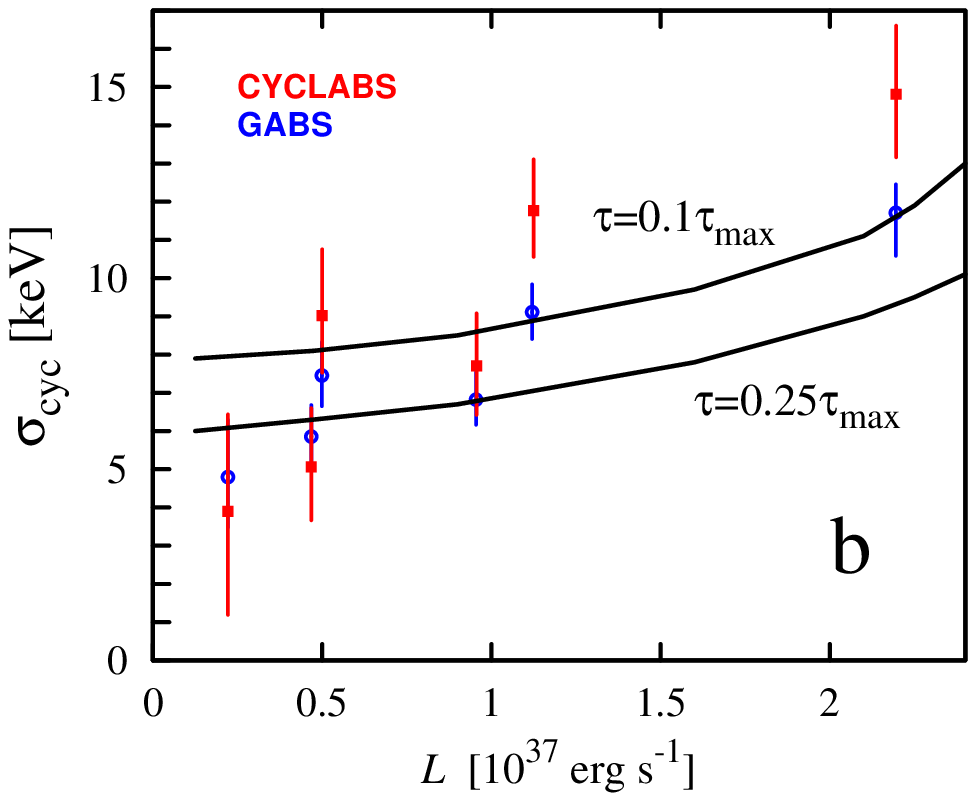}
\caption{(a) The cyclotron line energy and (b) the cyclotron line width
  dependencies on the 3--100~keV luminosity of GX 304--1. 
  Red squares show results obtained using the Lorentzian profile of the optical depth ({\sc cyclabs} model); 
  blue circles -- using the Gaussian profile ({\sc gabs} model) for the cyclotron absorption feature. 
  Luminosities are  calculated assuming the distance to the source of 2.4 kpc.
  Theoretical relations are given by black solid lines for the cyclotron energy at the NS surface for parameters 
  $E_{\rm cyc,0}=66.5\,{\rm keV}$, $i=70^\circ$,    $\alpha=10^\circ$, $L^*=2.7\times 10^{37}\,\ergs$. 
  A pure O-mode was considered. 
  The width of the resonance is measured at the level of 10 and 25 per cent of the maximum $\tau(E)$. 
}
\label{pic:corr}
\end{figure}

Three lower panels of Fig.\,\ref{pic:specs} show the ratio of spectra obtained at different orbits (see Table \ref{tab:observ}). 
We see that the spectra during brighter states are harder. 
Such a behaviour can be accounted for by the increasing Compton $y$-parameter with increasing mass accretion rate \citep{2015MNRAS.452.1601P}.
The most important feature in all spectral ratios is a wave-like structure at energies between 40 and 80 keV. 
We see an excess and depression relative to the general trend below and above 55 keV, respectively.

Our results confirm conclusions made by \citet{2012A&A...542L..28K} about the positive correlation of the cyclotron energy with flux in GX~304--1. 
Dependence of the cyclotron line centre on the observed flux is shown in Fig.~\ref{pic:corr}(a) for both shapes of the optical depth profile,  Lorentzian ({\sc cyclabs} model; red points) and Gaussian ({\sc gabs} model; blue points). 
The corresponding dependences of the line width on flux are shown in Fig. \ref{pic:corr}(b). 
We see that both the line energy and its width correlate with the flux. 
However, it is very important to keep in mind that the opacity profile differs significantly from a simple Gaussian or Lorentzian (see Sect.~{\ref{sec:TheMainFeatures}) at any intensity levels. 
These deviations leads to a difficulty with interpretation of the cyclotron energy obtained directly from the fits  using such simple phenomenological models.

\subsection{Observations versus theory}

For given geometrical parameters $i$, $\alpha$ (see Section \ref{sec:TheOpacityCalc}) and accretion luminosity $L$ (or corresponding final velocity $\beta_0$) we get the averaged opacity as a function of energy (see Fig.\,\ref{pic:opas_v}). 
Then the cyclotron line centroid energy can be estimated as 
\be
E_{\rm cyc}(L)\simeq \frac{\int E\,\tau(E,L)\d E}{\int \tau(E,L)\d E}.
\ee
For the particular set of parameters ($i=70^\circ$, $\alpha=10^\circ$, $L^{*}=2.7\times 10^{37}\,\ergs$ and surface cyclotron energy $E_{\rm cyc,0}=66.5\,{\rm keV}$) the theoretical dependence of the line centroid energy on luminosity is shown in Fig.\,\ref{pic:corr}(a). 
It is qualitatively consistent with the observed variations of the cyclotron line centroid with luminosity.

\begin{figure}
\centering 
\includegraphics[width=8.5cm]{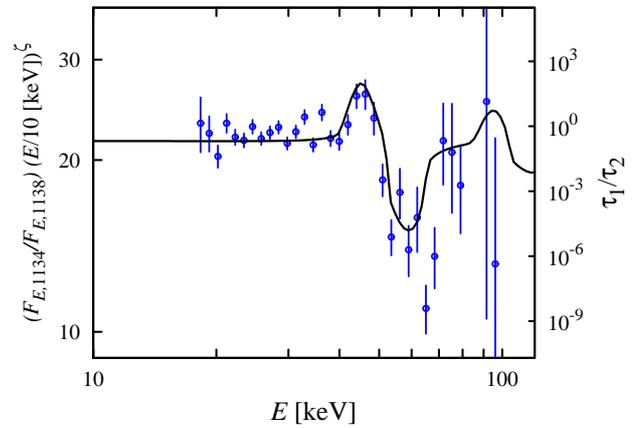} 
\caption{Ratio of spectra of GX 304--1 obtained in revolutions 1134 to 1138 corrected for different hardness ratio is given by blue circles (see also Fig.\,\ref{pic:specs}) and the ratio of the optical thicknesses for the case of pure O-mode polarization (solid line) 
for parameters $E_{\rm cyc,0}=66.5\,{\rm keV}$, $\zeta=0.55$, $i=70^\circ$, $\alpha=10^\circ$.}
\label{pic:opac_comp}
\end{figure}

Because there is no model of the line formation, we cannot predict well the shape of the absorption feature and its width. 
The line width in the observed spectrum can be different from the width of the resonance in optical thickness. 
We have measured the full width of the theoretical $\tau(E)$ dependence at the level of 10 and 25 per cent of the maximal $\tau$
as a function of the accretion luminosity (see   black solid lines in Fig.\,\ref{pic:corr}b).
We see that the model predicts a positive correlation, which is consistent qualitatively with the data. 
Moreover, we see that the observed ratios of the spectra obtained in different intensity states show very peculiar behaviour at the energies close to the cyclotron energy: the ratios reach local maximum below the cyclotron energy, then decrease and reach local minimum above the cyclotron energy (see Fig.\,\ref{pic:specs}). 
The same behaviour is seen in the ratios of the opacity profiles (see Fig.\,\ref{pic:opas_ration}(b) and Fig.\,\ref{pic:opac_comp}) calculated for different velocity gradients and, therefore, different accretion luminosity. 
This observation is consistent with our model that the cyclotron line forms by scattering of the hard radiation from the hotspots off the electrons in the accretion flow and that the behaviour of the cyclotron line is determined by the changes of the velocity profile in the line-forming region. 

In order to estimate the depth of the absorption feature, a detailed model of the line formation region is required. 
Here instead we make base our analysis on simple atmosphere models. 
According to the Schwarzschild-Schuster model \citep{1978stat.book.....M,1985cta..book.....S}, where the continuum and the absorption features form separately, the flux in the vicinity of the scattering feature can be estimated as follows
\be
F_E\propto \frac{1}{1+\tau_E},
\ee
where $\tau_E$ is the optical thickness of the line-forming region for a given photon energy $E$. 
The Thomson optical thickness of the line-forming region is $\sim$1 at $L\sim L^*$, and therefore the resonant optical thickness of the region is much larger than unity. 
As a result, the absorption features obtained with the Schwarzschild-Schuster model are extremely deep, which is not the case according to the observational data (see Fig.\,\ref{pic:specs}(a) and Table\,\ref{tab:observ}). 
This discrepancy might be caused by the increasing flux towards the photosphere which is related to
the energy-release inside the line-forming region where the accretion flow decelerates \citep{1982SvAL....8..330L}. 
Unfortunately, the exact energy deposition rate can be obtained only from a self-consistent model that includes solution of coupled  hydrodynamic and radiation transfer equations.
The photon redistribution over the energy can be also important. 
According to the Eddington model, where the regions of continuum and absorption feature formation are not separated, the absorption feature depth in a case of monochromatic scattering is $r_{E}\equiv F_{E}/F_{{\rm cont},E}\sim(\tau_{{\rm cont},E}/\tau_{E})^{1/2}$ \citep{1978stat.book.....M,1985cta..book.....S}, where $F_{E}$ and $F_{{\rm cont},E}$ are fluxes at a given energy $E$ with and without the absorption feature, and $\tau_{{\rm cont},E}$ is the optical thickness without resonant scattering. 
It gives $r_{E}\sim 0.03$ in our case. 
But if one takes into account  photon redistribution over the energy, then $r_E\sim (a\,\tau_{{\rm cont},E}/\tau_{E})^{1/4}$ \citep{1985cta..book.....S}, where $a=\Gamma_{\rm rad}/(2\omega_B v_{\rm th}/c)$ is the natural line width  parameter, $\Gamma_{\rm rad}$ is the cyclotron decay rate, $\omega_{B}$ is the cyclotron frequency and $v_{\rm th}=(2k_{\rm B}T/m_{\rm e})^{1/2}$. $\Gamma_{\rm rad}/\omega_B\sim 10^{-3}$ for $B=6\times 10^{12}\,{\rm G}$ \citep{Her1982} and therefore $a\sim 10^{-2}$. 
As a result, the expected absorption feature depth $r_{E}\sim 0.1$. 
This  is not far from the observed value (see Fig.~\ref{pic:specs}).

In order to compare theoretical and observational wave-like structures, we use the ratio of spectra obtained at revolution numbers 1134 and 1138, when the accretion luminosities differ by a factor of $10$, and correct the result for the difference in the spectral slope by multiplying  it by the power-law $E^\zeta$ with $\zeta=0.55$ (see Fig.\,\ref{pic:opac_comp}). 
For calculation of the optical thickness ratio we used $E_{\rm cyc,0}=66.5\,{\rm keV}$ and velocities on the bottom of line-forming region $\beta_0=0.1$ and $\beta_0=0.4$ for revolution numbers 1134 and 1138, respectively. 
During revolution 1134, the luminosity of GX 304--1 is very close to its critical value $L^*\approx (2\div 3)\times 10^{37}\,\ergs$ \citep{2015MNRAS.447.1847M}, while during revolution number 1138 it is far below the critical luminosity. 
This fact explains our choice of $\beta_0$ for the considered data sets (see also eq.\,(\ref{eq:L_general})). 
The theoretical opacities were averaged over the pulse period with parameters $i=70^\circ$ and $\alpha=10^\circ$ (see eq.\,(\ref{eq:tau_a1a2})). 
We see that theoretical ratio of the opacities shows qualitatively similar wave-like structure seen in the data.

The cyclotron energy $E_{\rm cyc,0}=66.5\,{\rm keV}$ is higher than line centroid energies measured from the data at any sub-critical luminosity states (see Fig.\,\ref{pic:corr} and Table\,\ref{tab:observ}).
This is well explained by our model: we always see a  red-shifted line absorption feature and only at critical luminosity the red-shift (due to the Doppler effect) turns to zero, but the estimated line centroid energy might be still red-shifted due to the complicated shape of the line. 
At sufficiently low mass accretion rate corresponding to luminosities below $10^{-3}L^*$ even the resonant optical thickness drops below unity. 
At this point the cyclotron line caused by scattering in the accretion flow disappears and the cyclotron line from the NS surface at $E\simeq E_{\rm cyc,0}$ can be observed in the spectrum.  
We thus expect that the positive correlation  between the line centroid energy and the accretion luminosity discussed in the present paper should be replaced by a negative correlation at sufficiently low luminosities.
Here we do not discuss the line formation in a region where the plasma flow is stopped by Coulomb collisions  and assume that the initial spectrum does not contain scattering features. 
In that sense, the theory does not give a complete model for the line formation, however, it points at the effects that have to be taken into account and  gives a qualitatively satisfactory description of the current observational data.

\section{Summary}
\label{discussion}

We have shown that the cyclotron line absorption feature in the spectra of sub-critical XRPs can be produced by scattering off moving electrons in the accretion channel. 
The variations of the line position with the accretion luminosity is naturally explained by changes of the plasma velocity profile near the NS surface, which is affected by the radiation flux from the hotspot. 
The presented model has several characteristic features: 
\begin{enumerate} 
\item the line centroid energy is positively correlated with the luminosity, 
\item the line centroid energy is always redshifted from the true cyclotron energy corresponding to the surface magnetic field at the NS pole, 
\item the cyclotron line width is also positively correlated with the luminosity, 
\item the line is expected to be asymmetric with excess at its red wing and it can hardly be approximated by widely used Lorentzian or Gaussian profiles, 
\item the phase-resolved cyclotron line centroid energy and the width are expected to be negatively and positively correlated with the pulse intensity, respectively. 
\end{enumerate}
In order to get the exact shape of the cyclotron line a self-consistent model of the line-forming region including radiation transfer and hydrodynamics is required,  which is beyond the scope of the present work.
Our theory is shown to be in good qualitative agreement with the available observational data on the sub-critical XRP GX 304--1 and has profound implications for the interpretation of the data on the cyclotron lines observed in XRPs.

\section*{Acknowledgments}

This work was supported by the University of Turku Collegium (SST), 
the Magnus Ehrnrooth Foundation (AAM),
the Academy of Finland grant 268740 (JP) and 
the German research Foundation (DFG) grant WE 1312/48-1 (VFS).
The authors are grateful to E.~M.~Churazov for developing the methods of analysis of INTEGRAL/IBIS data and providing the software.
The research used the data obtained from the HEASARC Online Service provided by the NASA/GSFC.


\bsp	
\label{lastpage}
\end{document}